\documentclass[a4paper,11pt]{article}

\usepackage{jheppub}
\usepackage{bm}
\usepackage{amsmath,amssymb,bm}
\usepackage{graphicx}

\usepackage{subfigure}

\newcommand{\be}{\begin{equation}}
\newcommand{\ee}{\end{equation}}
\newcommand{\bea}{\begin{eqnarray}}
\newcommand{\eea}{\end{eqnarray}}

\newcommand{\eq}[1]{Eq.~(\ref{eq:#1})}
\newcommand{\sect}[1]{Sec.~\ref{sec:#1}}
\newcommand{\appen}[1]{Appendix~\ref{sec:#1}}
\newcommand{\fig}[1]{Fig.~\ref{fig:#1}}

\newcommand{\del}{\partial}
\newcommand{\bra}{\langle}
\newcommand{\ket}{\rangle}

\newcommand{\stimulated}{\cite{Bao:2011pa}}
\newcommand{\Ts}{\mathcal{T}}
\newcommand{\Tco}{T_{c,0}}
\newcommand{\Ai}{\Vec{\bmA}}
\newcommand{\vevA}{\bra\Ai^2\ket}
\newcommand{\oscA}{\bm{\calA}_2}

\newcommand{\slow}{\varphi_s}
\newcommand{\fast}{\varphi_f}

\bmdefine{\bmA}{ \bm{A} }
\bmdefine{\bmD}{ \bm{D} }
\bmdefine{\bmF}{ \bm{F} }
\bmdefine{\bmgamma}{ \bm{\gamma} }
\newcommand{\calA}{ \mathcal{A} }
\newcommand{\calB}{ \mathcal{B} }
\newcommand{\calC}{ \mathcal{C} }
\newcommand{\calH}{ \mathcal{H} }
\newcommand{\calL}{ \mathcal{L} }
\newcommand{\calM}{ \mathcal{M} }
\newcommand{\calO}{ \mathcal{O} }
\newcommand{\calV}{ \mathcal{V} }
\newcommand{\vecv}{ \Vec{v} }
\newcommand{\vecx}{ \Vec{x} }
\newcommand{\hd}{ \Hat{d} }
\newcommand{\hA}{ \Hat{A} }
\newcommand{\hD}{ \Hat{D} }
\newcommand{\hcL}{ \Hat{\calL} }
\newcommand{\hbA}{ \Hat{\bmA} }
\newcommand{\tilvarphi}{ \Tilde{\varphi} }
\newcommand{\tbdy}{ \text{bdy} }
\newcommand{\Exp}[1]{\left\langle~#1~\right\rangle}
\newcommand{\head}[1]{%
\begin{flushleft}
        \underline{{\bf #1}}
\end{flushleft}
}
\title{The enhanced holographic superconductor: is it possible?}

\author[a]{Makoto Natsuume}
\author[b]{Takashi Okamura}

\affiliation[a]{KEK Theory Center, \\
Institute of Particle and Nuclear Studies, \\
High Energy Accelerator Research Organization,\\
Tsukuba, Ibaraki, 305-0801, Japan}
\affiliation[b]{Department of Physics, \\
Kwansei Gakuin University, \\
Sanda, Hyogo, 669-1337, Japan}

\emailAdd{makoto.natsuume@kek.jp}
\emailAdd{tokamura@kwansei.ac.jp}

\arxivnumber{1307.6875}
\preprint{KEK-TH-1648} 

\abstract{
It is known that time-dependent perturbations can enhance superconductivity and increase the critical temperature. If this phenomenon happens to high-$T_c$ superconductors, one could obtain room-temperature superconductors, but this is still an open issue experimentally. Meanwhile, we would like to understand this phenomenon from gravity dual and see if the enhancement is possible for holographic superconductors. Previous work (arXiv:1104.4098 [hep-th]) has studied this issue by adding a ``time-dependent chemical potential," but their analysis is questionable as a true dynamic equilibrium. In particular, the AdS boundary does not supply energy to the bulk spacetime in their setup. A more appropriate way to discuss the enhancement is to add a time-dependent vector potential, {\it i.e.}, a time-dependent electric field. However, the enhancement does not occur for holographic superconductors.
}


\begin{document}

\maketitle
\flushbottom

\section{Introduction and summary}

The AdS/CFT duality \cite{Maldacena:1997re,Witten:1998qj,Witten:1998zw,Gubser:1998bc} has many real-world applications such as QCD, quark-gluon plasma, and condensed-matter physics. Although basic nonequilibrium properties have been investigated using AdS/CFT, in real experiments, in particular in condensed-matter physics, one can control sources in various ways. Such situations have been investigated only recently in AdS/CFT. Examples include quenches and nonequilibrium phase transitions \cite{Bao:2011pa,Basu:2011ft,Bhaseen:2012gg,Gao:2012aw,Basu:2012gg,Nakamura:2012ae}. In this paper, we study holographic superconductors under time-dependent perturbations.  

It is known that time-dependent perturbations can enhance superconductivity and increase the critical temperature $T_c$ (see, {\it e.g.}, Refs.~\cite{review1,review2} for reviews). This is somewhat counterintuitive; since the source supplies energy to the system, the source would heat up the system and could destabilize the superconducting state. 

When the temperature is increased, more and more quasiparticles are excited, which block states for Cooper pair formation. Ultimately, this process leads to the disappearance of superconductivity at the critical temperature. Thus, the extraction of quasiparticles leads to the enhancement of superconductivity. Time-dependent perturbations effectively have this effect (Eliashberg theory). The perturbations excite quasiparticles near the Fermi surface to higher energy, so the number of quasiparticles decreases near the Fermi surface although the total number remains constant. We review the Eliashberg theory in \appen{eliashberg}.

This phenomenon has a potential application to high-$T_c$ superconductors since it would be possible to realize long-sought room-temperature superconductors. So, this possibility has been investigated, but this is still an open issue (see, {\it e.g.}, Ref.~\cite{Kaiser} for a recent attempt.)

Meanwhile, we would like to understand this phenomenon from gravity dual 
(holographic superconductors \cite{Gubser:2008px,Hartnoll:2008vx,Gubser:2008zu,Hartnoll:2008kx}) 
and to see if the enhancement is possible for holographic superconductors. Our results are summarized as follows:
\begin{enumerate}

\item
The work of enhanced holographic superconductors was pioneered by Bao, Dong, Silverstein, and Torroba \stimulated. They saw the enhancement by adding a ``time-dependent chemical potential." But their setup is questionable as a true dynamic situation (see \sect{summary} for a summary of problems). In particular, (i) the AdS boundary does not supply energy to the bulk in their setup, and (ii) they have several problems related to the bulk gauge symmetry.

\item
A more appropriate way to discuss the enhancement is to add a time-dependent vector potential $\vec{A}(t)$, {\it i.e.}, a time-dependent electric field%
\footnote{While this paper was in preparation, there appeared a preprint \cite{Li:2013fhw} which also studies holographic superconductors under a time-dependent electric field. After this work was completed, we were informed by Eva Silverstein that they had also studied this case and found no enhancement (private communication).}. 
However, the enhancement does not occur for holographic superconductors in contrast to Ref.~\stimulated. 

\end{enumerate} 
The basic reason of no enhancement is simple. We consider an Einstein-Maxwell-complex scalar system. The Maxwell field contributes to the  effective mass squared for the scalar field $|\Psi|$ (dual to the order parameter) as 
\be
m_\text{eff}^2 = m^2 + \left\{ - (-g^{tt})A_t^2 + g^{ii}\vec{A}^2 \right\}~.
%
\ee
The scalar mass effectively becomes tachyonic at a low enough temperature due to the gauge potential $A_t$. Thus, the gauge potential $A_t$ tends to destabilize the $\Psi=0$ solution, which leads to the superconductivity with $\Psi\neq0$. On the other hand, $\vec{A}$ tends to stabilize the the $\Psi=0$ solution, so it works as the suppression, not the enhancement. However, the problem is more subtle in a time-dependent case. Because $\vec{A}=\vec{A}(t)$, various Fourier modes for $\Psi$ are coupled. As we will see, this effect tends to compensate the suppression, but this effect is not large enough to have the enhancement.

This work may be also useful to study time-dependent holographic superconductors in general. In particular, 
\begin{enumerate}

\item
We present an energy flow analysis which gives a condition in order for the AdS boundary to supply energy in the bulk spacetime (\sect{energy_flow}).

\item
We propose a gauge-invariant definition of chemical potential $\mu_\text{inv}$ when $A_u(t,u) \neq 0$, where $A_u$ is the radial component of the Maxwell field (\sect{leading}). 

\end{enumerate} 

Holographic superconductors under time-dependent perturbations have been studied previously, so we would like to emphasize the difference. We study the response of the order parameter $\bra \calO \ket$ under time-dependent electric fields. On the other hand, Refs.~\cite{Hartnoll:2008vx,Hartnoll:2008kx} study the response of the current (conductivity) under time-dependent electric fields, and other works typically study the response of $\bra \calO \ket$  under the time-dependent source $\Psi^{(-)}(t)$ \cite{Basu:2011ft,Bhaseen:2012gg,Gao:2012aw,Basu:2012gg}.

\section{Preliminaries}
\label{sec:Model}

\head{Action and background spacetime}
\vspace{-0.3truecm}

We consider the $(p+2)$-dimensional $s$-wave holographic superconductors given by
\begin{align}
   S
  &= \int d^{p+2}x\, \sqrt{-g} \left[ R - 2 \Lambda 
  - \frac{1}{e^2} \left\{ 
  \frac{1}{4} F_{MN}^2 + \left\vert D \Psi \right\vert^2 + V\left( |\, \Psi\, |^2 \right) 
  \right\}
  \right]~,
\label{eq:full_action}
\end{align}
where 
%
\begin{align}
  & F_{MN} = 2\, \partial_{[M} A_{N]}~,
& & D_M := \nabla_M - i A_M~,
\label{eq:def-covariant_deri} \\
  & \Lambda
  = - \frac{p (p + 1)}{2\, L^2}~,
& & V= m^2 |\Psi|^2~.
\label{eq:def-potential_V}
\end{align}
We use Greek indices $\mu, \nu, \ldots$ for the $(p+1)$-dimensional boundary coordinates and use capital Latin indices $M, N, \ldots$ for the $(p+2)$-dimensional bulk spacetime coordinates $(x^\mu, u)$, where $u$ is the AdS radial coordinate. 

\begin{figure}[tb]
\centering
\scalebox{0.75}{ \includegraphics{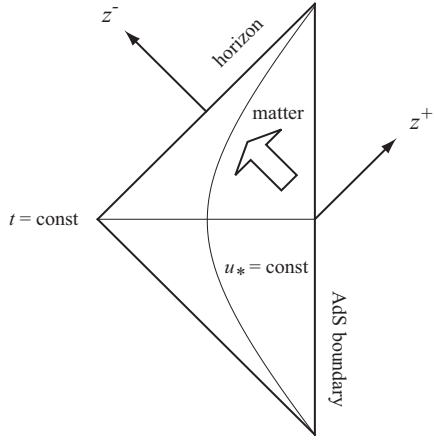} }
\vskip2mm
\caption{
The Eddington-Finkelstein coordinates.
}
\label{fig:light_cone}
\end{figure}

We take the probe limit $e\gg1$, where the backreaction of the matter fields onto the geometry can be ignored. When $p=2$, the background metric is given by the Schwarzschild-AdS$_4$ black hole:
\begin{subequations}
\label{eq:bg}
\begin{align}
ds_4^2 &= \left(\frac{r}{L}\right)^2 (-f(r)dt^2+ d\vecx_2^2) + L^2 \frac{dr^2}{r^2 f(r)}~, &
f(r) &= 1-\left(\frac{r_0}{r}\right)^3~,
\\
  &= \left( \frac{L\Ts}{u} \right)^2 \left(
  - f(u)\, dt^2 + d\vecx_2^2 + \frac{du^2}{\Ts^2\, f(u)} \right)~, &
  f(u) &= 1-u^3~,  &
  \Ts &:= \frac{4 \pi\, T}{3}~,
\label{eq:bg_metric-I} 
%
\end{align}
\end{subequations}
where $T$ is the Hawking temperature. It is also convenient to introduce the tortoise coordinate $u_*$ and light-cone coordinates $z^\pm$, where
\begin{align}
u_* &:= -\frac{1}{\Ts} \int^u \frac{du}{f}~, \\
z^\pm &:= t \pm u_*~.
%
\end{align}
Then, the $(t,r)$-part of the metric is given by
\begin{subequations}
\begin{align}
ds_2^2 &=  \left( \frac{L\Ts}{u} \right)^2 f (-dt^2 + du_*^2) \\
&= \left( \frac{L\Ts}{u} \right)^2 f (-dz^+{}^2 + 2dz^+du_*) \\
&= - \left( \frac{L\Ts}{u} \right)^2 f dz^+dz^-~. 
%
\end{align}
\end{subequations}
The coordinate $z^+$ is the ``ingoing" Eddington-Finkelstein coordinate, 
and $z^-$ is the ``outgoing" Eddington-Finkelstein coordinate (\fig{light_cone}). 
The horizon is located at a constant $z^-$.

\head{Equations of motion}
\vspace{-0.3truecm}
%
The equations of motion for matter fields are given by
\begin{subequations}
\label{eq:def-full_eq}
\begin{align}
  & D^2 \Psi - m^2 \Psi = 0~,
\label{eq:Psi_eom} \\
  & \nabla_N\, F^{MN} = j^M~,
\label{eq:def-EM_eq} \\
  & j^M
  = - i\, \left\{ \Psi^\dagger (D^M \Psi)
    - (D^M \Psi)^\dagger \Psi \right\}
  = 2\,\Im\left[\, \Psi^\dagger (D^M \Psi)\, \right]
  ~.
\label{eq:def-j}
\end{align}
\end{subequations}

Let us write $\Psi = |\, \Psi\, |\, e^{i \theta}$ and decompose the $\Psi$ equations of motion into the real and imaginary parts. It is convenient to use the gauge-invariant variable $\hA_M$ defined by
\be
\hA_M := A_M - \nabla_M \theta~.
\label{eq:def-hA}
\ee
Then, 
\be
  D_M \Psi
  = e^{i \theta} \left( \nabla_M - i\,\hA_M \right) |\, \Psi\, |
  =: e^{i \theta}~\hD_M |\, \Psi\, |
  ~.
\label{eq:covariant_deri-polar} 
\ee
Then, \eq{Psi_eom} is decomposed as 
\begin{subequations}
\begin{alignat}{2}
  & \text{Real:}  & 
  \nabla^2 |\, \Psi\, | - \left( m^2 + \hA_M^2 \right)~|\, \Psi\, | &= 0~,
\label{eq:Psi_eom_Re} \\ 
  & \text{Imaginary:}  & 
  - \frac{i}{ |\, \Psi\, | }~\nabla_M \left( |\, \Psi\, |^2\, \hA^M \right) &= 0~.
\label{eq:Psi_eom_Im}
\end{alignat}
\end{subequations}
Physically, the imaginary part represents the bulk current conservation equation $\nabla_M j^M = 0$ since
\be
j^M = 2\,\Im\left[\, \Psi^\dagger (D^M \Psi)\, \right] = - 2\,|\, \Psi\, |^2\, \hA^M~.
%
\ee
This fact will become important when we reexamine Ref.~\stimulated.

\head{Boundary conditions}
\vspace{-0.3truecm}

We impose the following boundary conditions:

\begin{itemize}

\item The AdS boundary: the asymptotic behavior of matter fields is given by 
\begin{subequations}
\label{eq:Psi-asympt}
\begin{align}
   \Psi(x, u)
  &\sim \exp\left( i\,\int^u_0 du' A_u(x, u') \right)
  \left\{ \Psi^{(-)}(x)\, \left( \frac{u}{L \Ts} \right)^{\Delta_-}
    + \Psi^{(+)}(x)\, \left( \frac{u}{L \Ts} \right)^{\Delta_+} \right\}
\label{eq:Psi-asym_behavior} \\
   \Delta_\pm
  &:= \frac{p + 1}{2}
  \pm \sqrt{ \left( \frac{p + 1}{2} \right)^2 + L^2\, m^2 }~,
\label{eq:def-Delta-Psi} \\
  A_\mu(x, u)
  & \sim A_\mu^{(0)}(x)
  + A_\mu^{(1)}(x)\, \left( \frac{u}{L \Ts} \right)^{p-1}~.
\label{eq:A-asym_behavior}
\end{align}
\end{subequations}
Equation~\eqref{eq:Psi-asym_behavior} is the expression in the $A_u\neq0$ gauge.
The fast falloff $ A_\mu^{(1)}$ represents the boundary current $\bra J^\mu \ket$, and the slow falloff $ A_\mu^{(0)}$ represents its source (external chemical potential $\mu$ and vector potential%
\footnote{As pointed out in \sect{leading}, this holds in the gauge $A_u=0$.}).
Similarly%
\footnote{For simplicity, we do not consider the ``alternative quantization," where the role of $\Psi^{(+)}$ and $\Psi^{(-)}$ is exchanged \cite{Klebanov:1999tb}.},
$\Psi^{(+)}$ represents the operator expectation value $\bra\calO\ket$, and 
$\Psi^{(-)}$ represents the external source for $\calO$.
We are interested in the spontaneous condensate, so we set $\Psi^{(-)}=0$.

\item Horizon: we impose the finiteness of the Euclidean action or finiteness of the energy-momentum tensor. The energy momentum tensor is given by
\be
   e^2 T_{MN} 
  = F_{ML} F_N{}^L
  + 2\, (D_{(M} \Psi)^\dagger\, (D_{N)} \Psi)
  + g_{MN} \left( - \frac{F^2}{4} - \big|\, D \Psi\, \big|^2
    - m^2 |\, \Psi\, |^2 \right)~.
\label{eq:energy_momentum}
\ee
Thus, we require that 
\be
\left|\, D \Psi\, \right|^2~, \quad
|\, \Psi\, |^2~,\quad
\left|\, F^2\, \right|\, \Big|_{u=1} < \infty~,
\label{eq:bc1}
\ee
or
\begin{subequations}
\label{eq:bc2}
\begin{align}
& (D_{(+} \Psi)^\dagger\, (D_{-)} \Psi)  \Big|_{u=1} = 0~,
\label{eq:bc-A_t-at_H} \\
& |\, F_{+-}\, |\, \Big|_{u=1} = 0~, \quad
|\, F_{+i}F_{-i}\, |\, \Big|_{u=1} = 0~, \quad
|\, F_{ij}\, |\, \Big|_{u=1} < \infty~.
\label{eq:bc-EM-at_H}
\end{align}
\end{subequations}
Equation~\eqref{eq:bc-A_t-at_H} is the incoming/outgoing wave boundary condition. In the gauge $A_u=0$ and $A_t(u=1)=0$, the condition reduces to $\del_{(+} \Psi^\dagger \del_{-)} \Psi |_{u=1} = 0$. We take the incoming wave boundary condition, so $\del_-\Psi|_{u=1}=0$. 
Similarly, for homogeneous perturbations, the condition 
$ |F_{+i}F_{-i} |\, |_{u=1} = 0$ reduces to $\del_+ A_i \del_- A_i |_{u=1} = 0$, the incoming/outgoing wave boundary condition for $A_i$.

\end{itemize}

For simplicity, we consider homogeneous solutions $A_M = A_M(t,u)$ and $\Psi = \Psi(t,u)$.
Also, we mainly consider 
$p=2$ and $L^2 m^2 = -2$. Then, $(\Delta_+,\Delta_-)=(2,1)$.

\section{The Case of the Missing Energy Flow}\label{sec:energy_flow}


\begin{figure}[tb]
\centering
\scalebox{0.75}{ \includegraphics{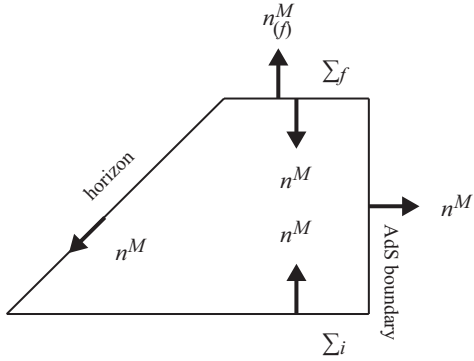} }
\vskip2mm
\caption{
The normal vectors for each hypersurfaces.
}
\label{fig:surface_element}
\end{figure}%

We would like to add a time-dependent source in a field theory. This corresponds to adding a time-dependent source on the AdS boundary. The supplied energy then flows into the bulk spacetime. Eventually, the energy is absorbed by the horizon and a dynamic equilibrium is achieved.

We would like to consider a time-dependent electromagnetic source. The source is given by the slow falloff, {\it e.g.}, $A_t(u) \sim \mu+\cdots$. So, one would replace the constant chemical potential by a time-dependent one: $A_t(t,u) \sim \mu(t)+\cdots$. This is the approach taken in Ref.~\stimulated. However, this has a serious problem. There is no energy flow from the AdS boundary to the bulk as we will see. 

For any Killing vector $\xi^N$, $\nabla_M (T^{MN} \xi_N) = 0$ from the conservation equation $\nabla_N T^{MN} = 0$. Thus, 
\be
\int_{\Sigma_i + \Sigma_f + \calH^+ + \tbdy} d\Sigma_M~T^{MN}\, \xi_N = 0
~,
\label{eq:Gauss}
\ee
where $\Sigma_i$ and $\Sigma_f$ are the initial and final spacelike hypersurfaces, respectively, extending from the AdS boundary to the future horizon $\calH^+$. 

The directed surface element $d\Sigma_M$ is given by $d\Sigma_M = n_M d\Sigma$, where $n_M$ is the ``outward-pointing" unit normal one-form. 
Note that the unit normal vector $n^M$ is inward-pointing for spacelike hypersurfaces (\fig{surface_element}). 
Let $n_{(f)}^M$ be the future-directed unit normal vector. The directed surface element then becomes
\begin{alignat}{2}
d\Sigma_M &= - d\Sigma~n^{(f)}_M &\quad& \text{for}~\Sigma_f~, \\
d\Sigma_M &= + d\Sigma~n^{(f)}_M &\quad& \text{for}~\Sigma_i~.
%
\end{alignat}

Then, \eq{Gauss} gives a conservation law. If one chooses the timelike Killing vector $\xi^M = (\partial_t)^M$, one has the energy conservation law:
\begin{subequations}
\label{eq:conservation_law-all}
\begin{align}
   \Delta E &:= E(\Sigma_f) - E(\Sigma_i) 
\nonumber \\
  &=  \int_{\tbdy} d\Sigma_M~T^{MN}\, \xi_N 
  + \int_{\calH^+} d\Sigma_M~T^{MN}\, \xi_N
  ~,
\label{eq:conservation_law} \\
   E (\Sigma)
  &:= \int_{\Sigma} d\Sigma~n^{(f)}_M\, T^{MN}\, \xi_N
  ~.
\label{eq:def-Q_xi}
\end{align}
\end{subequations}
On the AdS boundary, $n^u= -1/\sqrt{g_{uu}}$, so
\be
\int_{\tbdy} d\Sigma_M~T^{MN}\, \xi_N 
= - \int_{\tbdy} \dfrac{d\Sigma}{ \sqrt{g_{uu}} }~T_{ut}~. 
%
\ee
When the null generator of the future horizon is parametrized by $z^+$,
\be
\int_{\calH^+} d\Sigma_M~T^{MN}\, \xi_N 
= - \int_{\calH^+} dz^+\, d^px~T_{+t}~. 
%
\ee

Let us estimate the energy flow from the AdS boundary. On the AdS boundary, 
\be
T_{ut} \rightarrow \frac{1}{e^2} \left\{F_{ui}F_t^{~u} + 2(D_{(u} \Psi)^\dagger\, (D_{t)} \Psi) \right\}~,
%
\ee
so
\be
- \int_{\tbdy} \dfrac{d\Sigma}{ \sqrt{g_{uu}} }~T_{ut} 
= - \frac{1}{e^2} \int_{\tbdy} \frac{d\Sigma}{ \sqrt{g_{uu}} } 
\left\{ 
F_{ui} F_t^{~i} + 2 (D_{(u} \Psi)^\dagger\, (D_{t)} \Psi)
\right\}~.
\label{eq:energy_flow_bdy1}
\ee
When $\Psi$ has no source term ($\Psi^{(-)}=0$), the second term decays as $u^{2\Delta_+ -p-1}$, so it vanishes.

Let us rewrite \eq{energy_flow_bdy1} by boundary variables. According to the AdS/CFT duality, the boundary current expectation value is given by
\be
\bra J^\mu \ket = \left. \frac{1}{e^2} \sqrt{-g}\, F^{u\mu} \right|_{u=0}~,
%
\ee
and the boundary electric field is given by $-E^{(0)}_i = F^{(0)}_{ti} = \del_t A^{(0)}_i$. Thus,
\be
- \int_{\tbdy} \dfrac{d\Sigma}{ \sqrt{g_{uu}} }~T_{ut} 
= \int_{\tbdy} d^{p+1}x~\bra J^i \ket~E^{(0)}_i~.
\label{eq:energy_flow_bdy2}
\ee
Then the interpretation is clear:
\begin{itemize}
\item
From the boundary point of view, the energy flow from the boundary is just the Joule heat $\bra J^i \ket~E^{(0)}_i$. In other words, the boundary supplies energy in the form of the Joule heat.
\item
From the bulk point of view, $F_{ui}$ is a bulk magnetic field, so the expression is nothing but the Poynting vector $\vec{E}\times\vec{B}$ pointing the radial direction.
\end{itemize}
In order to supply the energy from the boundary, \eq{energy_flow_bdy2} must be nonvanishing, but $A_t(t,u)$ does not do the job. However, the analysis also indicates how one should include time-dependence; a bulk magnetic field $A_i(t,u)$ does the job. 

Let us consider the energy flow in the entire bulk spacetime. 
On $\calH^+$,
\begin{align}
T_{++} &\rightarrow \frac{1}{e^2} \left( F_{+ i}\, F_{+}{}^i + 2\, \big|\, D_+ \Psi\, \big|^2 \right)~, \\
T_{+-} &\rightarrow 0~,
%
\end{align}
where we used the boundary conditions \eqref{eq:bc1} and \eqref{eq:bc2}. 

Thus,
\begin{align}
\Delta E 
  &= - \frac{1}{e^2} \int_{\calH^+} dz^+\, d^px~
  \left( 2\, \big|\, D_+ \Psi\, \big|^2 + F_{+ i}\, F_{+}{}^i \right)
\nonumber \\
  &\quad + \int_{\tbdy} dt\, d^px~\bra J^{i} \ket~E^{(0)}_i
  - \frac{2}{e^2}\, \int_{\tbdy} \dfrac{d\Sigma}{ \sqrt{g_{uu}} }~
    (D_{(u} \Psi)^\dagger\, (D_{t)} \Psi)~.
\label{eq:E-conservation_law} 
\end{align}
This expression is valid irrespective of whether the system is in low temperature phase or not. 
Again the last term has no contribution when there is no source term $\Psi^{(-)}$. When $A_i=0$, the second term and the third term vanish as well.

If one has only $A_t(t,u)$, the boundary does not supply energy as discussed above. But the first line of \eq{E-conservation_law} represents the absorbed energy by the horizon (if one imposes the incoming-wave boundary condition.) The energy absorption occurs in the form of quasinormal modes. Since energy is absorbed by the horizon, the scalar field simply decays to a constant \cite{Maeda:2009wv}, so there is no interesting dynamic equilibrium.

This is different from Ref.~\stimulated. In fact, Ref.~\stimulated\, does not impose the incoming-wave boundary condition but imposes the regularity condition on the {\it static} $\Psi$. They try to justify their approach by focusing on the high frequency region and approximate the time-dependent solution by its average%
\footnote{Compare their analysis with our similar one in \sect{high_Omega}. In the high-frequency limit, we decompose the scalar field as the sum of the slow mode and the fast mode. It turns out that the source does not supply energy to the slow mode. So, the slow mode reduces to the quasinormal mode problem which is natural as we saw above. After taking a long-time average, the slow mode reduces to a static problem. On the other hand, the source supplies energy to the fast mode, and the fast mode is time-dependent. But the fast mode does not contribute the $\bra\calO\ket$.
}. 
This is how they obtain an equilibrium. However, this does not seem a true dynamic equilibrium we want since the boundary does not supply energy and the horizon does not absorb energy. Below we add a time-dependent vector potential $A_i(z^+)$ to study the enhancement in holographic superconductors. 

We are not saying that time-dependent gauge potential $A_t(t,u)$ is meaningless. We are saying that  $A_t(t,u)$ cannot be used to exchange energy between the AdS boundary and the bulk. But this is not the only way to supply energy to the bulk. In principle, one can supply energy to the bulk {\it directly} instead of supplying energy from the boundary. This becomes possible, {\it e.g.}, by a nonzero energy configuration of bulk matter fields. Our analysis does not consider such a case. 

For example, one can consider nonzero energy configuration for $\Psi$ on the bulk initial hypersurface $\Sigma_i$ and may consider its time evolution. In such a case, $A_t$ becomes time-dependent as well in general. Or one can consider $A_t(t,u)$ by coupling to the  {\it external bulk source.} However, from the AdS/CFT point of view, it is natural to consider a Dirichlet problem on the AdS boundary and natural to supply energy from the AdS boundary.

\section{Examination of previous work}

We have seen from the energy flow analysis that a time-dependent gauge potential $A_t(t,u)$ cannot supply energy from the boundary to the bulk. But Ref.~\stimulated\ has some other problems, and we would like to discuss them in this section.
Near the critical point, one can expand matter fields as a series in $\epsilon$, where $\epsilon$ is the deviation parameter from the critical point:
\begin{subequations}
\begin{align}
  & \Psi = \epsilon^{1/2} \left( \psi_1 + \epsilon\, \psi_2 + \cdots \right)
  ~,
\label{eq:pert_exp-Psi} \\
  & A_M = \bmA_M + \epsilon\, A_{1,M} + \epsilon^2\, A_{2,M} + \cdots
  ~,
\label{eq:pert_exp-A}
\end{align}
\end{subequations}
where boldface letters indicate background values. The other problems of Ref.~\stimulated\ are related to this perturbative expansion. The leading order equation is given by
\begin{align}
  & \nabla_N\, \bmF^{MN} =0~,
\label{eq:pert_zeroth-EM_eq}
\end{align}
and the first order equations are given by
\begin{subequations}
\label{eq:pert_1st-full_eq}
\begin{align}
   & \nabla_N\, F_1^{MN} = j_1^M~, \quad
   j_1^M = 2\, \Im\left[\, \psi_1^\dagger (\bmD^M \psi_1)\, \right]
  ~,
\label{eq:pert_1st-EM_eq} \\
  & \bmD^2 \psi_1 - m^2 \psi_1 = 0~.
\label{eq:pert_1st-Psi_eq-I}
\end{align}
\end{subequations}
Our main analysis from \sect{enhanced_H^3} will not add a time-dependent $A_t(t,u)$ but add $A_i(t,u)$, so readers who are interested in our case can skip this section 
and go directly to \sect{enhanced_H^3}.

\subsection{Leading order}\label{sec:leading}

The $t$ and $u$ components of the leading order equation \eqref{eq:pert_zeroth-EM_eq} become $\del_u \bmF_{tu} = \del_t \bmF_{tu} =0$. Thus the solution is given by $\bmF_{tu} = \mu_0$ or 
\begin{subequations}
\label{eq:pert_zeroth-unitary-homo-comp-massless}
\begin{align}
  \bmA_t(t, u) &= \mu_0 (1-u) + \del_t \bmgamma(t, u)~, \\
  \bmA_u(t,u) &= \del_u \bmgamma(t, u)~.
\end{align}
Reference~\stimulated\, introduces a time-dependence by choosing 
\begin{align}
\bmgamma(t, u) = (1-u) \int dt'~(\mu(t') - \mu_0)~.
\label{eq:silverstein}
\end{align}
\end{subequations}
Note that $A_u\neq0$ for their choice although we normally choose the $A_u=0$ gauge.

Equations~\eqref{eq:pert_zeroth-unitary-homo-comp-massless} look like a time-dependent configuration. However, the bulk gauge field has the gauge symmetry
\begin{subequations}
\be
A_M(x,u) \to A_M(x,u) - \del_M \Lambda(x,u)~,
\label{eq:bulk_gauge}
\ee
or
\begin{align}
A_\mu(x,u) &\to A_\mu(x,u) - \del_\mu \Lambda(x,u)~, \\
A_u(x,u) &\to A_u(x,u) - \del_u \Lambda(x,u)~.
%
\end{align}
\end{subequations}
Then, the deformation \eqref{eq:silverstein} is gauge-equivalent to the time-independent case by choosing $\Lambda=\bmgamma$. Since the deformation is just a gauge degree of freedom, it does not seem a meaningful one%
\footnote{After this work was completed, we learned that Simeon Hellerman made a similar point during a seminar by one of the authors of Ref.~\stimulated\ (private communication).}. 



A related issue is that their definition of the chemical potential $\mu_\text{naive} := A_t(x, u=0)$ is not gauge invariant under \eq{bulk_gauge}. What is the definition of the chemical potential when $A_u(t,u) \neq 0$?
One way is to transform to the $A_u=0$ gauge, but then the deformation \eqref{eq:pert_zeroth-unitary-homo-comp-massless} reduces to the time-independent one as we saw, so their chemical potential is constant. Another way is to define a gauge-invariant chemical potential. We propose 
\begin{align}
\mu_\text{inv}
&:= \int^0_1 du~F_{ut}(x, u) 
  \label{eq:def-mu_2} \\
&= A_t(x, u=0) - A_t(x, u=1) - \partial_t \int^0_1 du~A_u(x, u)
\end{align}
as a gauge-invariant chemical potential%
\footnote{After this work was completed, we learned that Refs.~\cite{Nakamura:2006xk,Nakamura:2007nx} made a similar proposal in the context of static situations.}.
The solution \eqref{eq:pert_zeroth-unitary-homo-comp-massless} gives $\mu_\text{naive}=\mu(t)$, but $\mu_\text{inv}=\mu_0$. 

In the gauge $A_u=0$, $\mu_\text{inv}$ reduces
\begin{align}
    & \mu_\text{inv} \to A_t(x,u=0) - A_t(x,u=1)
    ~.
  \label{eq:def-mu_3}
\end{align}
In the gauge $A_u=0$, the bulk gauge symmetry is not completely fixed. The gauge transformation of the form $\Lambda=\Lambda(x)$ is allowed, and $A_t(x,u)$ transforms as $A_t(x,u) \to A_t(x,u) - \del_t\Lambda(x)$. This gives the gauge symmetry of the dual field theory in the sense that one transforms the external source $A_\mu(u=1)$ \cite{Maeda:2010br}.
One can fix the gauge for $A_t$ by requiring $A_t(x,u=1)=0$. Then, $\mu_\text{inv}$ becomes
\be
\mu_\text{inv} \to A_t(x,u=0)~.
%
\ee
Thus, $\mu_\text{inv}$ reduces to the naive definition of the chemical potential, but it differs when $A_u(t,u) \neq 0$.

\subsection{First-order}\label{sec:current_conserv}

When Ref.~\stimulated\, solves the $\psi_1$ equation at the first order, they impose the unitary gauge/London gauge, where the phase $\theta$ of $\Psi$ vanishes. However, a care is necessary to impose a gauge condition on $\Psi$ even though we expand around $\Psi=0$. At leading order, this cannot fix a gauge (since $\Psi=0$ at leading order.) At leading order, the deformation \eqref{eq:silverstein} itself is a gauge choice. But then, it is too restrictive to impose an additional gauge choice such as the unitary gauge. Since $A_M$ and $\theta$ appear in the combination of $A_M - \nabla_M\theta$, one should not impose conditions both on $A_M$ and $\theta$. If one chooses $A_M$, $\theta$ must be determined by solving an equation of motion.

This problem becomes apparent if one considers the imaginary part of $\psi_1$ equation of motion \eqref{eq:Psi_eom_Im}. Reference~\stimulated\, solves $|\psi_1|$ equation of motion in the unitary gauge. Even in the unitary gauge, the imaginary part is not absent, but they are not taking the equation into account. As mentioned in \sect{Model}, the imaginary part represents the current conservation and is nontrivial:
\be
\nabla_M j_1^M 
\propto -\del_t(|\psi_1|^2 \hbA_t) + f u^2 \del_u\left(\frac{\Ts^2 f}{u^2}|\psi_1|^2 \hbA_u \right) = 0~.
%
\ee
Note that $\hbA_M=\bmA_M$ in the ``unitary gauge." Since $\bmA_t$ and $\bmA_u$ are determined at leading order and $|\psi_1|$ are determined at first order, the equation gives a nontrivial condition. Put differently, one cannot choose the background gauge field $\bmgamma(t,u)$ freely.

\subsection{Summary of problems}\label{sec:summary}

We point out that Ref.~\stimulated\, has a number of problems. We summarize them here for convenience:
\begin{enumerate}

\item {\it No energy flow from the AdS boundary to the bulk} (\sect{energy_flow}): 
In order for the AdS boundary to supply energy to the bulk, $A_t(t,u)$ is not sufficient, and a bulk magnetic field $A_i(t,u)$ is necessary.

\item {\it No energy flow from the bulk to the horizon} (\sect{energy_flow}): 
Since the AdS boundary does not supply energy, $\Psi$ simply decays to a constant if one imposes the incoming-wave boundary condition on the horizon. They impose the static regularity condition to avoid the problem.

\item {\it Problem of gauge choice} (Sec.~\ref{sec:leading}, \ref{sec:current_conserv}): 
Their time-dependence is just a gauge choice. Their definition of the chemical potential is not gauge invariant. Although they make a gauge choice by choosing $\bmA_t$ and $\bmA_u$, they impose an additional gauge choice (unitary gauge), which leads to the next problem.

\item {\it Lack of conservation equation} (\sect{current_conserv}): 
Even in the unitary gauge, the imaginary part of $\Psi$ equation of motion is not absent, but the equation, which is the current conservation equation, is not taken into account. Since $\bmA_t$ and $\bmA_u$ are determined at leading order and $|\psi_1|$ are determined at first order, the equation gives a nontrivial condition, which is unclear if it is satisfied. 

\end{enumerate}
They expand around $\Psi=0$ even though they take the unitary gauge; Problems~3 and 4 are related to this. In principle, one can avoid these problems if one solves the full matter equations without using the perturbative expansion. In the low-temperature phase where $\Psi \neq 0$, one can then impose the unitary gauge, and the Maxwell field would become $F_{tu}\neq \mu_0$ because of the coupling to $\Psi$, which is clearly different from the static solution $F_{tu} = \mu_0$. 

However, even if one avoids Problems~3 and 4 in this way, one cannot avoid Problems~1 and 2 which are independent from the perturbative expansion problem. Their system is rather limited as a dynamical system. One can also see this from their degrees of freedom. Their system consists of $A_t(t,u), A_u(t,u)$, and $\Psi(t,u)$. When $\Psi=0$, the system is just a $(1+1)$-dimensional electromagnetic field, so there should be no dynamical degree of freedom. Therefore, $\Psi\neq 0$ is necessary for such a system to be dynamical, and there should be no phase where $\Psi$ always vanishes. Since their system is a rather limited dynamical system, one probably had better study a different setup (such as the next section one).

\section{No enhanced holographic superconductor}\label{sec:enhanced_H^3}

\subsection{Equations of motion}\label{sec:enhanced_eom}

Our energy flow analysis shows that a time-dependent gauge potential $\bmA_t(t,u)$ cannot supply energy from the boundary to the bulk and one needs $\Ai(t,u)$. However, $\Ai$ does not seem very useful to enhance superconductivity. This is because the gauge field contributes to the effective mass squared for $|\Psi|$ as 
\be
m_\text{eff}^2 = m^2 + \left\{ - (-g^{tt})\bmA_t^2 + g^{ii}\Ai^2 \right\}
%
\ee
from \eq{Psi_eom_Re}. Thus, $\bmA_t$ tends to destabilize the normal state, which leads to the superconductivity. On the other hand, $\Ai$ tends to stabilize the normal state. Namely, $\Ai$ works as the suppression of the superconductivity and there is no enhancement. 
In fact, a large enough magnetic field destroys the superconducting state (See, {\it e.g.}, Refs.~\cite{Hartnoll:2008kx,Nakano:2008xc,Albash:2008eh,Albash:2009ix,Montull:2009fe,Maeda:2009vf} for holographic realizations.) 

However, the problem is more subtle in a time-dependent case. Because $\Ai=\Ai(t,u)$, various Fourier modes for $\psi_1$ are no longer decoupled. To see more explicitly, decompose the vector potential term into the time-averaged part $\vevA$ and the oscillatory part $\calA_2(t)$: 
\be
m_\text{eff}^2 = m^2 + \left\{ - (-g^{tt})\bmA_t^2 + g^{ii} \left( \vevA + \calA_2(t) \right) \right\}~.
%
\ee
The time-averaged part works as the suppression as described above, and various Fourier modes for $\psi_1$ couple through $\calA_2(t)$. The resulting equation of motion is difficult to analyze analytically, and one generically needs a numerical computation. But we can analyze it analytically both in the low-frequency limit and in the high-frequency limit. 

In the low-frequency limit, we will show that one can diagonalize the Fourier-transformed equations of motion and that the time-dependent part $\calA_2(t)$ tends to compensate the effect of $\vevA$. This behavior is indeed very similar to the Eliashberg theory (\appen{eliashberg}): the enhancement is possible only after the enhancement term compensates the suppression term by $\vevA$. However, the effect of $\calA_2(t)$ is never larger than the effect of $\vevA$ for holographic superconductors. So, the critical temperature is lower than the original critical temperature $\Tco$ with $\Ai=0$, and the enhancement does not occur. 

As in Ref.~\stimulated, we solve the matter equations of motion perturbatively. In the presence of $\Ai$, the leading order equation $\nabla_N \bmF^{iN} =0$ becomes
\be
  0
  = - \del_t^2 \Ai + \Ts f \del_u(\Ts f \del_u \Ai) 
  = - \del_t^2 \Ai + \del_{u_*}^2 \Ai 
  = - 4 \del_+\del_- \Ai~.
\label{eq:pert_zeroth-EM_eq-i-homo-II}
\ee
With the incoming-wave boundary condition, we choose the solution
\be
\Ai = \Ai(z^+)~.
%
\ee
Thus, our choice of the time-dependent gauge field is
\begin{subequations}
\begin{align}
\text{Ours:} &&
\bmA_t &= \mu_0 ( 1 - u )~,&
\bmA_u &= 0~,&
\Ai &= \Ai(z^+)~,&
\label{eq:pert_zeroth-sol-p2} \\
\text{Ref.~\stimulated:} &&
\bmA_t &= \mu(t)(1-u)~,&
\bmA_u &= - \! \int \! dt' (\mu(t')-\mu_0)~,&
\Ai &= 0~.&
\end{align}
\end{subequations}
%
%
%
%
In particular, our $\bmA_t$ is standard, and our gauge choice $\bmA_u = 0$ is standard as well. 
Below we consider the vector potential $\Ai$ given by
\begin{align}
  & \big|\, \Vec{\bmA}(z^+)\, \big|
  = \frac{\delta E}{\Omega}\, \sin\left( \Omega\, z^+ \right)
  ~.
\label{eq:def-sinusoidal_pert}
\end{align}
The applied boundary electric field is then given by $|\Vec{E}^{(0)}(t)| = \delta E \cos (\Omega t)$. 

\head{Real space expression}

At first order, the scalar equation of motion \eqref{eq:pert_1st-Psi_eq-I} in $(z^+,u_*)$-coordinates becomes
\label{eq:pert_1st-Psi_eq-homo-staticG-p2}
\be
\left[\, 
  - 2 \left( \partial_{u_*} + i\bmA_t \right) \partial_+
  - \partial_{u_*}^2 
  + f\, \left\{ 
     \Ts^2\, \frac{ L^2 m^2 \!+\! 2 \!+\! u^3 }{u^2}
     - \frac{1}{f} \bmA_t^2
     + \Ai^{\, 2}
  \right\}
\, \right]\, \varphi = 0
~,
\label{eq:pert_1st-Psi_eq-homo-staticG-p2-III} 
\ee
where we choose the combination
\be
\varphi(u,z^+) := \frac{\psi_1}{u}
\label{eq:def-varphi}
\ee
to eliminate a term which is linear in $\del_{u_*}$. We will set $L^2m^2=-2$ to simplify the equation. We also decompose the vector potential term into the time-averaged part $\vevA$ and the oscillatory part $\oscA(z^+)$: 
\begin{subequations}
\begin{align}
\frac{1}{\Ts^2} \Ai^{\, 2} &=: \oscA(z^+) + \frac{1}{\Ts^2} \vevA~,
\\
\bra f(z^+) \ket &:= \lim_{t \to \infty}\, \frac{1}{2 t} \int^t_{-t} dz^+~f(z^+)~.
\label{eq:def-Exp_f}
\end{align}
\end{subequations}
For the vector potential \eqref{eq:def-sinusoidal_pert}, 
\be
\frac{1}{\Ts^2} \vevA = 2 \left( \frac{\delta E}{2\Ts\Omega} \right)^2 =: 2\calC^2~, \qquad
\oscA(z^+) = - 2\calC^2 \cos( 2\Omega z^+ )~.
\label{eq:sinusoidal_decompose}
\ee
Then, in the coordinate $u$, the equation of motion is rewritten as
\begin{subequations}
\begin{align}
& \left[\, 
  \frac{2}{\Ts}\, \left( \partial_u - i\,\bm{\calA}_t \right) \partial_+
  - \partial_u \left( f\, \partial_u \right) 
  +\left( u - f \bm{\calA}_t^2  + \frac{1}{\Ts^2}\, \vevA \right)
  + \oscA
\, \right]\, \varphi = 0
~, \\
& \bm{\calA}_t(u) := \frac{\bmA_t}{\Ts f}~.
%
\end{align}
\end{subequations}
We write this equation as 
\be
\left[\, \frac{2}{\Ts}\,\hd_u\, \del_+ + \hcL_u + \oscA(z^+)\, \right]\, \varphi(u,z^+) = 0
~,
\label{eq:pert_1st-Psi_eq-homo-staticG-p2-arrangeI-EOM}
\ee
where 
\begin{subequations}
\begin{align}
\hd_u &:= \partial_u - i\, \bm{\calA}_t~,
\label{eq:def-hd_u} \\
\hcL_u &:= - \partial_u \left( f\, \partial_u \right) + \calV(u)
~,
\label{eq:def-hcL_u} \\
\calV(u) &:= u - f \bm{\calA}_t^2 + \frac{1}{\Ts^2}\, \vevA~.
\label{eq:def-calV}
\end{align}
\end{subequations}

\head{Momentum space expression}

Let us write \eq{pert_1st-Psi_eq-homo-staticG-p2-arrangeI-EOM} in momentum space. We make the Fourier transformation in $z^+$, {\it e.g.},
\begin{subequations}
\label{eq:def-FT}
\begin{align}
  & \tilvarphi(u, \omega)
  = \int^\infty_{-\infty} \frac{dz^+}{ \sqrt{2\, \pi} }~
  e^{i \omega z^+}\, \varphi(u, z^+)
  ~,
\label{eq:def-FT-varphi} \\
  &\tilde\oscA(\omega)
  = \int^\infty_{-\infty} \frac{dz^+}{ \sqrt{2\, \pi} }~
  e^{i \omega z^+}\, \bm{\calA}_2(z^+)
  ~.
\label{eq:def-inv-FT-A2}
\end{align}
\end{subequations}
For our vector potential \eqref{eq:sinusoidal_decompose},
\begin{align}
\tilde\oscA(\omega)
&= -\sqrt{2 \pi}~\calC^2\,
  \left\{ \delta(\omega + 2\, \Omega) + \delta(\omega - 2\, \Omega) \right\}~.
\label{eq:sinusoidal_pert} 
%
\end{align}
From \eq{sinusoidal_pert}, the Fourier mode $\tilvarphi(u,\omega)$ couples to the modes with $\omega + 2\Omega$ and $\omega - 2\Omega$. But the mode with $\omega + 2\Omega$ couple to the ones with $\omega + 4\Omega$ and $\omega$, and so on. Consequently, modes whose $\omega$ differs by $2n\Omega$ ($n$: integer) all couple. In fact, after the Fourier transformation, \eq{pert_1st-Psi_eq-homo-staticG-p2-arrangeI-EOM} becomes
\begin{subequations}
\begin{align}
& \left[\, 
  \hcL_u 
  - \frac{2\, i\, (\omega+2n\Omega)}{\Ts}\,  \hd_u
\, \right]\, \tilvarphi^{(\omega)}_n
  = \calC^2\, \left\{ \tilvarphi^{(\omega)}_{n-1}
  + \tilvarphi^{(\omega)}_{n+1}
  \right\}
  ~,
\label{eq:sinusoidal_pert-Psi_eq-homo-staticG-p2-EOM-FT-I} \\
& \tilvarphi^{(\omega)}_n(u)
  := \tilvarphi(u, \omega + 2\, n\, \Omega)~, \quad
  \left( - 2\, \Omega < \omega < 2\, \Omega \right)~.
\label{eq:def-tilvarphi_n} 
%
\end{align}
\end{subequations}

Our problem reduces to coupled differential equations for a chosen $\omega$. We are interested in adding time-dependence to the static system, so we focus on $\omega=0$: 
\begin{subequations}
\label{eq:sinusoidal_pert-Psi_eq-homo-staticG-p2-EOM-FT}
\begin{align}
\hcL_u \tilvarphi^{}_n
&=
  \calC^2\, \left\{ \tilvarphi^{}_{n-1}
    + 2\, i\, n\, w\, \hd_u\, \tilvarphi^{}_n
    + \tilvarphi^{}_{n+1}
     \right\}
  ~,
\label{eq:sinusoidal_pert-Psi_eq-homo-staticG-p2-EOM-FT-II} \\
w &:= \dfrac{2\, \Omega}{ \calC^2\Ts }
  = \dfrac{8\, \Ts\, \Omega^3}{ \delta E^2 }
  ~.
\label{eq:def-w}
\end{align}
\end{subequations}
More explicitly,
\begin{subequations}
\label{eq:sinusoidal_pert-Psi_eq-homo-staticG-p2-EOM-FT-explicit}
\begin{align}
  & \cdots
\nonumber \\
  \hcL_u \tilvarphi^{}_{-1}
  &= \calC^2\, \left( \tilvarphi^{}_{-2}
    - 2\, i\, w\, \hd_u\, \tilvarphi^{}_{-1}
    + \tilvarphi^{}_{0} \right)
  ~,
\label{eq:sinusoidal_pert-Psi_eq-homo-staticG-p2-EOM-FT-explicit-mI} \\
  \hcL_u \tilvarphi^{}_0
  &= \calC^2\, \left( \tilvarphi^{}_{-1}
    + \tilvarphi^{}_{1} \right)
  ~,
\label{eq:sinusoidal_pert-Psi_eq-homo-staticG-p2-EOM-FT-explicit-0} \\
  \hcL_u \tilvarphi^{}_1
  &= \calC^2\, \left( \tilvarphi^{}_{0}
    + 2\, i\, w\, \hd_u\, \tilvarphi^{}_{1}
    + \tilvarphi^{}_{2} \right)
  ~,
\label{eq:sinusoidal_pert-Psi_eq-homo-staticG-p2-EOM-FT-explicit-pI} \\
  & \cdots
\nonumber
\end{align}
\end{subequations}
The boundary conditions for $\tilvarphi_n(u)$ are as follows:
\begin{enumerate}
\item
On the horizon, we imposed the incoming wave boundary condition for $\varphi(u,z^+)$, which becomes  the regularity condition for Fourier components $\tilvarphi_n(u)$. 
\item
The asymptotic behavior of $\tilvarphi_n(u)$ is given by $ \tilvarphi_n(u) \sim u^{\Delta_+-1} = u$ from \eq{def-varphi}.
\end{enumerate}

These coupled differential equations are rather difficult to handle analytically, but one can handle them analytically in two limits, the low-frequency limit and the high-frequency limit. We will see that there is no enhancement in these limits.

\subsection{Low-frequency limit}

We will specify our low-frequency limit below, but first we approximate \eq{sinusoidal_pert-Psi_eq-homo-staticG-p2-EOM-FT-II} to a system with a finite number of modes. As we saw, modes with all integer $n$ are coupled, but it is natural to assume that modes with large $n$ are not excited. So, consider a finite number of modes $n = 0, \pm 1, \cdots, \pm N$. We will go back and check the assumption. Then,
\eq{sinusoidal_pert-Psi_eq-homo-staticG-p2-EOM-FT-II} becomes
\begin{subequations}
\label{eq:sinusoidal_pert-Psi_eq-homo-staticG-p2-EOM-FT-matrix1}
\begin{align}
  & 
  \hcL_u 
  \Vec{\varphi}^{}
  = \calC^2 
  \calM_{2N+1} \Vec{\varphi}^{}
  ~,
\label{eq:sinusoidal_pert-Psi_eq-homo-staticG-p2-EOM-FT-matrix_form1} \\
  & \Vec{\varphi}^{}
  := {}^t\!  \begin{pmatrix} \tilvarphi^{}_{-N}
    & \tilvarphi^{}_{-(N-1)} & \cdots
    & \tilvarphi^{}_{N-1} & \tilvarphi^{}_{N}
    \end{pmatrix}
    ~,
\label{eq:def-vecvarphi} 
\end{align}
where $ \calM_{2N+1}$ is a $(2 N + 1) \times (2 N + 1)$ matrix given by
\be
%
  \calM_{2 N + 1}
  := \begin{pmatrix}
      \ddots & \vdots & \vdots & \vdots & \vdots & \vdots &  \\
      \cdots & -4iw\hd_u & 1 & 0 & 0 & 0 & \cdots \\
      \cdots & 1 & -2iw\hd_u & 1 & 0 & 0 & \cdots \\
      \cdots & 0 & 1 & 0 & 1 & 0 & \cdots \\
      \cdots & 0 & 0 & 1 & 2iw\hd_u & 1 & \cdots \\
      \cdots & 0 & 0 & 0 & 1 & 4iw\hd_u & \cdots \\
       & \vdots & \vdots & \vdots & \vdots & \vdots & \ddots \\
  \end{pmatrix}
  ~.
\label{eq:def-band_matrix}
\ee
\end{subequations}

As a low-frequency limit, we take the limit
\be
N w \simeq N \frac{\Omega}{\calC^2 \Ts} \simeq N \frac{\Omega^3 \Ts}{\delta E^2} \ll 1~.
%
\ee
In this limit, one can ignore the terms with the operator $\hat{d}_u$ in \eq{sinusoidal_pert-Psi_eq-homo-staticG-p2-EOM-FT-II}%
\footnote{More precisely, we assume $ |\, n\, w~\hd_u \tilvarphi_n\, | \ll |\, \tilvarphi_{n \pm 1}\, | $.}.
Then, $\calM_{2N+1}$ reduces to a tridiagonal matrix:
\begin{align}
  \calM_{2 N + 1}
  &\sim \begin{pmatrix}
      0 & 1 & 0 & \cdots \\
      1 & 0 & 1 & \cdots \\
      0 & 1 & 0 & \cdots \\
      \vdots & \vdots & \vdots & \ddots \\
  \end{pmatrix}~.
\label{eq:def-tridiagonal}
\end{align}

The tridiagonal matrix $\calM_{2N+1}$ is diagonalized as 
\be
\calM_{2N+1} \vecv_k = \lambda_k \vecv_k~, \quad
\left( k = 1, 2, \cdots, 2 N + 1 \right)~,
\label{eq:eigen_system-eqn} 
\ee
where the integer $k$ labels each eigenvalue and
\begin{subequations}
\label{eq:eigen_values-calM}
\begin{align}
  & \lambda_k = 2\, \cos\frac{\pi k}{2 (N + 1)}~,
\label{eq:lambda_k} \\
  & \vec{v}_k = \frac{1}{ \sqrt{N + 1} } {}^t\!  \displaystyle{ 
   \begin{pmatrix} 
        \sin\frac{k\pi}{2(N+1)},
    &  \sin\frac{2k\pi}{2(N+1)}, & \cdots,
    &  \sin\frac{(2N+1) k\pi}{2(N+1)} 
    \end{pmatrix} }
    ~,
\label{eq:eigenvector}
\end{align}
or the $n$th component $v_{k,n}$ is given by
\begin{align}
  & v_{k,n}
  = \frac{1}{ \sqrt{N + 1} }~\sin\left[ \frac{k \pi}{2} \left(\frac{n}{N+1} +1 \right)
  \right]~,
& & \left( n = - N, \cdots, -1, 0, 1, \cdots, N \right)
~.
\label{eq:eigenvector_comp}
\end{align}
\end{subequations}
Then, \eq{sinusoidal_pert-Psi_eq-homo-staticG-p2-EOM-FT-matrix_form1}
is diagonalized as
\begin{subequations}
\label{eq:sinusoidal_pert-Psi_eq-homo-staticG-p2-EOM-FT-matrix-D}
\begin{align}
  \hcL_u 
  \tilvarphi^{}_{\text{NM},k}
  & = \calC^2 \lambda_k \,
  \tilvarphi^{}_{\text{NM},k}
  ~,
\label{eq:sinusoidal_pert-Psi_eq-homo-staticG-p2-EOM-FT-matrix_form-D} \\
  \Vec{\varphi}^{}
  & =: \vec{v}_k\, \tilvarphi^{}_{\text{NM},k}
  ~.
\label{eq:def-normal_vecvarphi} 
\end{align}
\end{subequations}
Writing \eq{sinusoidal_pert-Psi_eq-homo-staticG-p2-EOM-FT-matrix_form-D} more explicitly, $\varphi^{}_{\text{NM},k}$ satisfies
\begin{align}
  & \left\{ - \partial_u \left( f\, \partial_u \right)
  + u - f\, \bm{\calA}_t^2
  + \frac{1}{\Ts^2} \vevA \left( 1 - \frac{1}{2}\, \lambda_k \right)
 \right\}\,
 \tilvarphi^{}_{\text{NM}, k}
 = 0
  ~.
\label{eq:sinusoidal_pert-Psi_eq-homo-staticG-p2-EOM-FT-matrix_form-D-exp}
\end{align}
Because the eigenvalue part contributes with the minus sign, the oscillatory part $\oscA$ indeed tends to compensate the effect of $\vevA$ which works as the suppression of superconductivity. The time-dependent electric field would enhance the instability of the system if an eigenvalue with  $\lambda_k > 2$ exists. However, $\lambda_k \leq 2$ from \eq{lambda_k}, so there is no mode which enhances the instability%
\footnote{
The largest eigenvalue is $\lambda_1$, so the $k=1$ mode gives the dominant contribution to the instability. 
The maximum is $\lambda_1=2$ when $N\to\infty$. From \eq{eigenvector_comp}, the components of the eigenvector $\vec{v}_1$ have the maximum value $O(N^{-1/2})$ at $n=0$ and decays as $O(N^{-3/2})$ at $n=O(N)$, so the modes with large $n$ are indeed hard to excite.}.

\subsection{High-frequency limit}\label{sec:high_Omega}

In order to consider the high-$\Omega$ limit, it is convenient to go back to the real space equation of motion \eqref{eq:pert_1st-Psi_eq-homo-staticG-p2-arrangeI-EOM}. We consider the situation where the time scale of the electric field $1/\Omega$ is much shorter than the time scale of the system. The following discussion is essentially the ``averaging method" described in Sect.~30 of Ref.~\cite{landau}. The time scale of the system is most likely to be of order of $T_c^{-1}$, but we do not really need to specify the time scale below. 

In such a situation, the system evolves with its time scale slowly and at the same time oscillates rapidly with $1/\Omega$. Thus, one expects that $\varphi$ is the sum of the slow mode $\slow$ and the fast mode $\fast$: 
\be
\varphi(u, z^+) = \slow(u, z^+) + \fast(u, z^+)~.
\label{eq:averaging}
\ee
The slow mode $\slow$ is almost constant over the period $2\pi/\Omega$ while the mean value of $\fast$ over the period should vanish%
\footnote{
In reality, the averaging does not provide a sharp high-frequency cutoff at $\Omega$, and even $\slow$ has some fast components. We discuss this issue in \appen{reexamine_high}.}. 
Denoting the time-average over $2\pi/\Omega$ as ``~$\bar{~}$~," $\bar{\varphi}_f = 0$. Substituting \eq{averaging} into \eq{pert_1st-Psi_eq-homo-staticG-p2-arrangeI-EOM}, one obtains
\be
 \left( \frac{2}{\Ts}\, \hd_u\, \partial_+
    + \hcL_u \right) \slow
  + \left( \frac{2}{\Ts}\, \hd_u\, \partial_+
    + \hcL_u \right) \fast
  + \oscA\, \slow
  + \oscA\, \fast = 0~.
\label{eq:averaging_eom}
\ee
However, so far we just introduced a set of redundant variables $(\slow, \fast)$. So, we have to specify the $\slow$ equation or the $\fast$ equation. This is done by requiring that $\slow$ satisfies the time-averaged equation of \eq{averaging_eom}. Then, the $\fast$ equation follows uniquely. The $\slow$ and $\fast$ equations are
\begin{subequations}
\label{eq:averaging_eom1}
\begin{align}
\left( \frac{2}{\Ts}\, \hd_u\, \partial_+ + \hcL_u \right) \slow(u, z^+)
&= 
  - \overline{ \oscA(z^+)\, \slow(u, z^+) }
  - \overline{ \oscA(z^+)\, \fast(u, z^+) }
  ~,
\label{eq:slow_eom1} \\
\left( \frac{2}{\Ts}\, \hd_u\, \partial_+ + \hcL_u \right) \fast(u, z^+)
&=
  - \Delta\left[\, \oscA(z^+)\, \slow(u, z^+)\, \right]
  - \Delta\left[\, \oscA(z^+)\, \fast(u, z^+)\, \right]
  ~,
\label{eq:fast_eom1} 
\end{align}
where
\be
\Delta f(z^+) := f(z^+) - \overline{ f(z^+) }~.
\label{eq:def-Delta}
\ee
\end{subequations}
Note that one can consistently set $\bar{\varphi}_f = 0$ from \eq{fast_eom1}.

So far we made no approximations. The equations are simplified via approximations. The slow mode $\slow$ is constant inside averaged expressions, so
\begin{subequations}
\begin{align}
\overline{ \oscA(z^+)\, \slow(u, z^+) } & \simeq \overline{\oscA(z^+)} \slow(u, z^+) \simeq 0~, 
 \label{eq:approx1} \\
\Delta\left[\, \oscA(z^+)\, \slow(u, z^+)\, \right] & \simeq  \oscA(z^+)\, \slow(u, z^+)~,
%
\end{align}
\end{subequations}
where we have used $\overline{\oscA(z^+)}=0$. In \eq{fast_eom1}, the term $\del_+ \fast$ is proportional to large $\Omega$, so it is kept, but the other terms with $\fast$ can be ignored%
\footnote{
More explicitly,
\begin{align*}
  & \frac{1}{\Ts}\, O\big( |\, \hd_u\, \partial_+ \fast\, | \big)
  = \frac{\Omega}{\Ts}\, O\big( |\, \hd_u\, \fast\, | \big)
  \gg O\big( |\, \hcL_u\, \fast\, | \big)~,
\\
\intertext{when $\Omega \gg \Ts$, and}
  & \frac{1}{\Ts}\, O\big( |\, \hd_u\, \partial_+ \fast\, | \big)
  = \frac{\Omega}{\Ts}\, O\big( |\, \hd_u\, \fast\, | \big)
  \gg O\big( |\, \oscA\, \fast\, | \big)
  ~,
%
\end{align*}
by assuming $\Omega/\Ts \gg O( |\, \oscA\, | )$.
}.
Then, we reach at
\begin{subequations}
\begin{align}
\left( \frac{2}{\Ts}\, \hd_u\, \partial_+ + \hcL_u \right) \slow
&\simeq 
  - \overline{ \oscA(z^+)\, \fast(u, z^+) }
  ~, \quad (\text{slow})
\label{eq:slow_eom2} \\
\frac{2}{\Ts}\, \hd_u\, \partial_+ \fast
&\simeq
  -  \oscA(z^+)\, \slow(u, z^+)
  ~. \quad (\text{fast})
\label{eq:fast_eom2} 
\end{align}
\end{subequations}
From \eq{fast_eom2}, the $\oscA\slow$ term is the source term for $\fast$. $\oscA$ is fast oscillating, so it is natural that $\fast$ is proportional to $\oscA$. From \eq{slow_eom2}, $\slow$ could depend on $\fast$ only through $\overline{\oscA\fast}$. 

The solution of \eq{fast_eom2} which is consistent with the boundary conditions is given by 
\begin{subequations}
\begin{align}
\fast(u, z^+)  &\simeq
 - \frac{\Ts}{2}\, e^{i \chi(u)}\, 
  \int^u_0 du'\, e^{- i \chi(u')}\,
  \int^{z^+} dv\, \oscA(v)\, \slow(u', v)~,
\label{eq:fast_sol} \\
   \chi(u) &:= \int^u_0 du'\, \bm{\calA}_t(u')~.
\label{eq:def-chi} 
\end{align}
\end{subequations}
%
%
Substitute \eq{fast_sol} into \eq{slow_eom2}. The source term $\overline{\oscA\fast}$ becomes
\begin{subequations}
\begin{align}
\overline{ \oscA(z^+)\, \fast(u, z^+) }
& \simeq - \frac{\Ts}{2}\, e^{i \chi(u)}\, 
  \int^u_0 du'\, e^{- i \chi(u')}\,
  \overline{ \oscA(z^+) \int^{z^+} dv\, \oscA(v)\, \slow(u', v) } \\
& \simeq - \frac{\Ts}{2}\, e^{i \chi(u)}\, 
\int^u_0 du'\, e^{- i \chi(u')}\, \slow(u', z^+)\,
\overline{ \oscA(z^+)\, \int^{z^+} dv\, \oscA(v)}\,
    \\
&\propto \overline{ \calB(z^+)\, \del_+\calB(z^+) }~,
\qquad  
\calB(z^+) :=  \int^{z^+} dv\, \oscA(v) \\
&= \frac{1}{2}\, \overline{ \del_+ \calB^2(z^+) }
= 0~.
\label{eq:eval_slow_source}
\end{align}
\end{subequations}
This conclusion holds for any periodic vector potential $\Ai(z^+)$.
%
%
Thus, \eq{slow_eom2} becomes 
\begin{align}
\left( \frac{2}{\Ts}\, \hd_u\, \partial_+ + \hcL_u \right)\, \slow(u, z^+)
\simeq 0~.
\label{eq:slow_eom3}
\end{align}
The $\fast$-dependence on $\slow$ is completely gone. This conclusion is not trivial. To reach this conclusion, it is essential that our equation of motion \eqref{eq:fast_eom2} is first order in ``time" $z^+$. This makes the source term $\overline{\oscA\fast}$ the time-average of a total derivative in \eq{eval_slow_source}. If the equation were second order in time, the conclusion would be different. Since the source does not supply energy to the slow mode, the slow mode problem reduces to the quasinormal mode problem with $\vevA$ as mentioned in \sect{energy_flow}.

The problem can be further simplified by taking the long-time average. Under the long-time average, 
\be
\Exp{ \partial_+ \slow }
  := \lim_{t \to \infty}\, \frac{1}{2 t}
  \int^t_{-t} dz^+~\partial_+ \slow(u, z^+)
  = \lim_{t \to \infty}\, \frac{\slow(u, t) - \slow(u, -t)}{2\, t}
  = 0~.
\label{eq:long_time}
\ee
Then, the slow mode equation \eqref{eq:slow_eom3} reduces to the static problem with $\vevA$:
\be
\hcL_u\, \Exp{ \slow(u, z^+) } 
\simeq 0~.
%
\ee

Thus, we obtain the following conclusions for $\slow$ and $\fast$:
\begin{itemize}

\item
For the slow mode $\slow$, the $\fast$-dependence is completely gone, and the problem formally reduces to the quasinormal mode problem with $\vevA$. Moreover, after the long-time average, the problem further reduces to the static problem with $\vevA$, and the naive argument at the beginning of \sect{enhanced_eom} applies. Namely, the vector potential works as the suppression of superconductivity, and there is no enhancement. Since $\slow$ has no $\oscA$ dependence, the supplied energy is absorbed by the fast mode.

\item
The fast mode $\fast$ behaves as $\fast \sim u^{\Delta_+}$ from \eq{fast_sol} under the asymptotic boundary condition $\varphi \sim u^{\Delta_+ -1}$; the fast mode has a faster falloff than the slow mode. Thus, the fast mode $\fast$ does not contribute to $\bra\calO\ket$, and $\bra\calO\ket$ is determined by the slow mode $\slow$ only. 

\end{itemize}
Note the difference between our analysis and the analysis of Ref.~\stimulated. Our problem reduces to the static problem like Ref.~\stimulated, but this is only for the slow mode. The fast mode is driven by the source and is time-dependent. However, the fast mode has a faster falloff so that the fast mode does not contribute to $\bra\calO\ket$.

\section{Discussion}\label{sec:discussion}

Although the previous analysis saw the enhancement \stimulated, we questioned if their setup is a true dynamic equilibrium. It is more appropriate to add a time-dependent electric field, but we find no enhancement in contrast to Ref.~\stimulated. Of course, our setup is different from theirs and cannot be directly compared, but our interpretation is that their enhancement comes from the improper analysis. 

We close with a list of relevant questions we have not mentioned:
\begin{enumerate}

\item {\it Is the enhancement possible in intermediate frequencies?}
In intermediate frequencies, we carried out a numerical computation. We see no enhancement although we have not explored the full parameter space. We numerically solved \eq{sinusoidal_pert-Psi_eq-homo-staticG-p2-EOM-FT-II} with small numbers of modes ({\it e.g.}, 3 or 5 modes corresponding to $N=1,2$) by a shooting technique. 

\item {\it Possibility to get the enhancement in other models.}
We saw that the enhancement is unlikely in our simple model, but the enhancement could be possible in a more complicated model. In the Eliashberg theory, the hierarchy of two scales is necessary to get the enhancement:
\be
\frac{1}{\tau} \ll T_c~,
\label{eq:fermi_liquid}
\ee
where $\tau$ is the (inelastic) relaxation time of the quasiparticle (See \appen{eliashberg}). Note that this is nothing but the condition for the Fermi liquid, so the Eliashberg theory relies on the weakly-coupled quasiparticle picture of the Fermi liquid. Our holographic superconductor is a Ginzburg-Landau-like theory, and it is not clear how to estimate $\tau$, but it is natural to expect that it is the order of temperature. Then, the lack of the hierarchy may be the reason of no enhancement. One probably needs to consider a model with the hierarchy of the two scales.

\item {\it Is a bulk fermion necessary to see the enhancement?}
Our holographic superconductor is an Einstein-Maxwell-scalar system, and there is no bulk fermion. 
From the Fermi liquid point of view, the hierarchy of scales \eqref{eq:fermi_liquid} comes from the fermionic nature of the quasiparticle, so this is an interesting direction to explore. The bulk fermion can describe the Fermi liquid as well as the non-Fermi liquid. However, it is not immediately obvious if our holographic analysis is insufficient. The Eliashberg theory essentially reduces to a Ginzburg-Landau theory. Moreover, as we saw in \sect{enhanced_H^3}, the oscillatory part $\oscA(t)$ indeed tends to compensate the effect of $\bra \Ai^{\, 2} \ket$. The critical temperature is lower than the original critical temperature $\Tco$ with $\Ai=0$ but is higher than the one with $\bra \Ai^{\, 2} \ket$ only. In this sense, we saw an ``enhancement."%
\footnote{Although $\bra \Ai^{\, 2} \ket =$constant for $p=2$, $\bra \Ai^{\, 2} \ket$ is a function of $u$ for $p \neq 2$. Then, the situation with $\bra \Ai^{\, 2} \ket$ corresponds to the one with a stationary current on the boundary since $\bra J_i\ket \propto \bmF_{ui} = \del_u \vec{\bmA}$.}
Our problem is that the oscillatory part is never larger than the time-average part.

\item {\it Beyond the probe limit.}
We take the probe limit in our analysis. In our case, the supplied energy is eventually absorbed by the black hole, so the black hole plays the role of an infinite heat bath. In the Eliashberg theory, what plays the role of the heat bath is the phonon system. Time-dependent perturbations excite quasiparticles to higher energy, but the quasiparticles are not stable. These quasiparticles decay to phonons, so the supplied energy is eventually absorbed by phonons or lattice. But in real systems, one does not have an infinite heat bath, so the heating effect of the phonon system is an important issue. One can incorporate this effect and can show that the enhancement breaks down for $\Omega \gtrsim T_c$. 

So, it is important to take the backreaction onto the geometry into account for holographic superconductors as well, but this effect does not help for the enhancement. Incidentally, this may be an important issue for Ref.~\stimulated. Reference~\stimulated\ claimed the enhancement for $\Omega \gtrsim T_c$. Their enhancement is likely to be a gauge artifact as we discussed in this paper. Even if their setup can be justified in some way, the backreaction onto the geometry may change their conclusion. 

\item {\it Exploring the full phase diagram.}
We expand matter fields at the critical point to simplify the analysis, so we cannot exclude the possibility of the enhancement far from the original critical point. In fact, in the Eliashberg theory, there are two branches of solutions, and the branch which smoothly connects to the original critical point is unstable (\appen{eliashberg}). The system actually has the first-order phase transition from the normal state to the stable branch. Our perturbative analysis is unlikely to see the stable branch if the holographic superconductor has the similar phase diagram as the Eliashberg theory (\fig{GL_E}). Thus, exploring the full phase diagram is an interesting issue to study. Although we focus on the unphysical branch, the enhancement itself occurs even in the unstable branch according to the Eliashberg theory.

\end{enumerate}

Finally, let us compare the Eliashberg theory and the observed enhancement for cuprates \cite{Kaiser}. It is not our purpose here to discuss the observed enhancement for cuprates, but the Eliashberg theory will be clearly inadequate to explain this phenomenon from the following reasons. First, Ref.~\cite{Kaiser} claimed that the observed data do not fit with the Eliashberg theory. 
Second, Ref.~\cite{Kaiser} observed the enhancement in the underdoped region, but this is the non-Fermi liquid phase. The Eliashberg theory is based on the BCS theory or the Fermi liquid picture, so there is {\it a priori} no reason to trust the Eliashberg theory. Moreover, their data points seem to overlap with the pseudogap region. Thus, the observed enhancement is likely to involve the yet mysterious pseudogap physics. However, clearly we need more experimental results to have a definite conclusion. 

We saw that the enhancement based on the Eliashberg mechanism is unlikely for holographic superconductors. If holographic superconductors resemble the cuprates in some way, our result may suggest that the observed enhancement is due to a completely different mechanism from the Eliashberg theory.


\acknowledgments
We would like to thank Elena Caceres, Xi Dong, Roberto Emparan, Jerome Gauntlett, Steven Gubser, Sean Hartnoll, Simeon Hellerman, Gary Horowitz, Shin Nakamura, Joe Polchinski, Tadakatsu Sakai, Shigeki Sugimoto, Hirotaka Yoshino, and Jan Zaanen for useful discussions. MN would also like to thank the Banff International Research Station for Mathematical Innovation and Discovery (BIRS) and its program ``Holography and Applied String Theory" (Feb.\ 10-15 2013) for the opportunity to present this work and for a stimulating environment. 
We would also like to thank the Yukawa Institute for Theoretical Physics at Kyoto University and its programs YKIS2012 (Oct.\ 15-19, 2012) and ``Gauge/Gravity Duality" (Sep.\ 24-Oct.\ 26, 2012) where part of this work was carried out, 
and the KIAS-YITP joint workshop 2013 (July 1-5, 2013) for the opportunity to present this work. 
TO would like to thank Takashi Oka and the workshop ``Mathematical Method and Mathematical Physics for the Materials Science'' (June 15-17, 2012) held in the RIKEN Nishina Center for Accelerator-Based Science; he pointed us Ref.~\cite{Kaiser} at the workshop, which motivated this work. 
The research of MN and TO were supported in part by a Grant-in-Aid for Scientific Research (23540326) from the Ministry of Education, Culture, Sports, Science and Technology, Japan. 

\appendix

\section{Review of Eliashberg theory}\label{sec:eliashberg}

In the Eliashberg theory, time-dependent perturbations excite quasiparticles near the Fermi surface to higher energy leaving room for Cooper pair formation near the Fermi surface. The quasiparticles are not stable and decay to phonons. So, the supplied energy is eventually absorbed by phonons or lattice, which plays the role of the heat bath.

The Eliashberg theory is summarized by a Ginzburg-Landau-like equation:
\be
\frac{7\zeta(3)}{8\pi^2} \left(\frac{\Delta}{\Tco}\right)^2
= (1-t) +f~.
\label{eq:GL_E}
\ee
Here, $\Delta$ is the condensate, $t:=T/\Tco$ with the equilibrium critical temperature $\Tco$ , and $f$ represents nonequilibrium terms: we will explain their explicit forms below. \eq{GL_E} is valid near the critical point $\Delta/\Tco \ll 1$. When $f$ vanishes, $\Delta \propto (1-t)^{1/2}$, which is the standard mean-field behavior. 

As time-dependent perturbations, we consider the time-dependent vector potential $\vec{A}(t)$ below. In this case, the nonequilibrium terms $f$ are given by
\be
f = 
\frac{1}{4} \tau\alpha\, \frac{\Omega}{\Tco} G\left(\frac{\Delta}{\Omega}\right)
- \frac{\pi}{2} \frac{\alpha}{\Tco} 
+ O\left( (\tau\alpha)^2 \right) + O\left( (\Omega/\Tco)^2 \right)~,
\label{eq:Eliashberg}
\ee
where $\tau$ is the (inelastic) relaxation time of the quasiparticle and
\begin{align}
\alpha &:= 2D A_\Omega A_{-\Omega}~, \\
\big| \vec{A} \big| &= A_\Omega \sin(\Omega t)~. 
%
\end{align}
Here, $D$ is the diffusion constant. The first term of \eq{Eliashberg} represents the effect of the {\it enhancement}, and the second term represents the effect of the time-averaged value of $\vec{A}^2$, which works as the {\it suppression}. One key feature which is important to our discussion is that a large $\tau$ is favorable to the enhancement since the enhanced term is proportional to $\tau$.

The function $G$ is given by
%
%
%
%
\be
G(u) = \begin{cases}
  \frac{2\pi u}{ \sqrt{1-u^2} }~, & (u<1/2)~, \\
  \frac{2}{ u+\frac{1}{2} } \left[ K(k) + 4u^2 \left\{ \Pi(\hat{\alpha}^2|k) - K(k) \right\} \right]~, & (u>1/2)~,
\end{cases}
%
\ee
where the functions $K(k)$ and $\Pi(\hat{\alpha}^2|k)$ are complete elliptic integrals of the first kind and the third kind, respectively, and 
\be
k := \frac{ u-\frac{1}{2} }{ u+\frac{1}{2} }~, \quad
\hat{\alpha}^2 := \frac{1}{ (2u+1)^2 }~.
%
\ee
The function $G(u)$ has the maximum $G(1/2) \approx 3.63$ at $u=1/2$.
Also, $G(u)$ can be approximated as
\be
G(u) \to \begin{cases}
  2 \pi u~, & (u \to 0)~, \\
  \frac{ 2\ln(2.9u) }{u} & (u \to\infty)~.
\end{cases}
%
\ee

\begin{figure}[tb]
\centering
\scalebox{0.9}{ 
\includegraphics{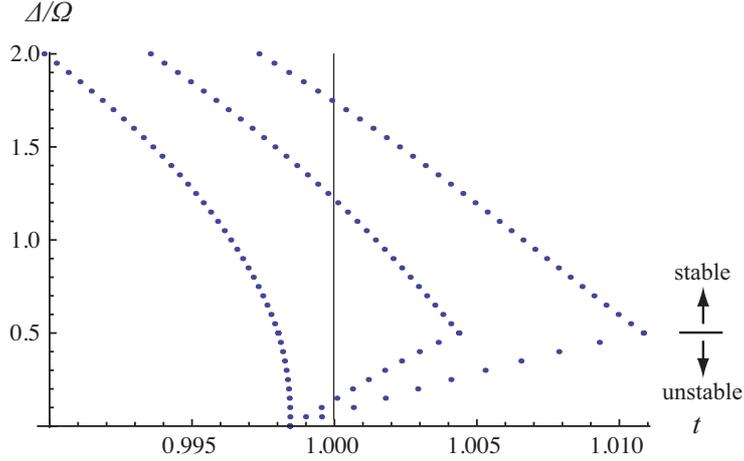} 
}
\vskip2mm
\caption{Solutions to \eq{GL_E} when $\Omega/\Tco =1/7$ and $\alpha/\Tco = 10^{-3}.$ $\tau\alpha = 10^{-3}, 0.05$, and 0.1 (from left to right).}
\label{fig:GL_E}
\end{figure}

The Eliashberg theory has five dimensionless parameters: $t, \Delta/\Tco, \Omega/\Tco, \alpha/\Tco$, and $\tau\alpha$. In \fig{GL_E}, numerical solutions to \eq{GL_E} are given for fixed $\Omega/\Tco, \alpha/\Tco$, and $\tau\alpha$. The plots show that a large $\tau$ is favorable to the enhancement; we make a simple estimate below.

Since $G(u)$ has different behaviors as we vary $u$, there are three interesting regions of $\Delta/\Omega$.

\head{(i) $\Delta/\Omega <1/2$}
In this region, $G(\Delta/\Omega)$ is approximately linear in $\Delta$, and the $O(\Delta^2)$ term in \eq{GL_E} can be ignored. Then, one gets
\be
\Delta \simeq \frac{2}{\pi}\frac{1}{\tau\alpha} \Tco (t-1) + \frac{1}{\tau}~.
%
\ee
Note that $\Delta \propto (t-1)$ instead of the standard mean-field behavior $\Delta \propto (1-t)^{1/2}$.

\head{(ii) $\Delta/\Omega >1/2$}
In this case, the $O(\Delta^2)$ term in \eq{GL_E} cannot be ignored. $G(u)$ falls as $1/u$ for a large $u$, so one has the standard mean-field behavior $\Delta \propto (1-t)^{1/2}$. As we will see, this region is most interesting physically, but we skip the detailed analysis since we focus on the region which smoothly connects to the original critical point in this paper for technical reasons. 

\head{(iii) $\Delta/\Omega =1/2$}
Since $G(u)$ has the maximum at $u=1/2$, this case gives the maximum temperature $T_\text{max}$. Substituting $\Delta/\Omega =1/2$ into \eq{GL_E}, one gets
\be
\frac{T_\text{max}}{\Tco} -1 \approx 
\frac{3.63}{4} \tau\alpha\, \frac{\Omega}{\Tco} 
- \frac{7\zeta(3)}{32\pi^2} \left(\frac{\Omega}{\Tco}\right)^2
- \frac{\pi}{2} \frac{\alpha}{\Tco} 
- 0.17 \tau\alpha\, \left(\frac{\Omega}{\Tco}\right)^2
\label{eq:Tmax}
\ee
($7\zeta(3)/(32\pi^2) \approx 0.0266$). The last term is the $O(\Omega^2)$ term which we did not write explicitly in \eq{Eliashberg}. This term also works as a suppression, so we would like this term not to make a considerable contribution. Comparing the second and the last term, the last term is negligible if $(\tau\alpha) \lesssim O(0.1)$.

Let us estimate the maximum  $T_\text{max}$ as we vary $\Omega$ (we denote it as $T_\text{max, max}$) . Solving $\del_\Omega T_\text{max} = 0$, $T_\text{max}$ has the maximum at $\Omega_o \approx 17.0 (\tau\alpha) \Tco + O((\tau\alpha)^2)$. Then,
\be
\left. \frac{T_\text{max, max}}{\Tco} -1 \right|_{\Omega_o} \approx 
- \frac{\pi}{2} \frac{\alpha}{\Tco} 
+ 7.73 (\tau\alpha)^2 + O\left( (\tau\alpha)^3 \right)~.
%
\ee
The second term can dominate over the first term if $\tau \gg 1/\Tco$. On the other hand, if $\tau =O(1/\Tco)$, it is difficult to get an enhancement.

As one deviates from $\Omega_o$, $T_\text{max}$ decreases, and the enhancement is hard to occur both at low and high frequencies. At low frequencies, the third term of \eq{Tmax} works as the suppression, so the enhancement is possible if the first term is larger than the third term:
\be
\frac{3.63}{4} \tau\alpha\, \frac{\Omega}{\Tco} > \frac{\pi}{2} \frac{\alpha}{\Tco}~,
%
\ee
which determines the minimum frequency $\Omega_\text{min}$ for the enhancement to occur:
\be
\Omega_\text{min} 
= \frac{2\pi}{3.63} \frac{1}{\tau}
\approx \frac{1.73}{\tau}~.
%
\ee

At high frequencies, the second term of \eq{Tmax}, which also works as the suppression, makes a considerable contribution, so there is the maximum frequency $\Omega_\text{max}$. But the heating effect below is more important to determine $\Omega_\text{max}$ in reality.

\head{Stability of states}

As one can see in \fig{GL_E}, there are two branches of solutions, $u<1/2$ and $u>1/2$ in a given temperature. In order to determine which solution is stable, one needs a non-equilibrium generalization of free energy. 

From such a free energy, one can show that the $u<1/2$ branch is {\it unstable} for all $T$ and the $u>1/2$ branch is globally {\it stable} for $\Tco < T <  T_\text{K} < T_\text{max}$. Thus, the system actually has the first-order phase transition from the normal state $\Delta=0$ to the $u>1/2$ solution at $T_\text{K}$; this $T_\text{K}$ is the true critical temperature.

\head{Heating and the maximum frequency}

We studied holographic superconductors in the probe limit, so the black hole plays the role of an infinite heat bath. In real systems, one does not have an infinite heat bath, and the phonon system experiences the heating effect, namely the phonon system deviates from the equilibrium. Naturally, this occurs for $\Omega \gtrsim \Tco$. One can incorporate this effect and can show that the enhancement indeed breaks down for $\Omega \gtrsim \Tco$, so there is a maximum frequency $\Omega_\text{max}=O(\Tco)$ for enhancement.

\section{A more careful analysis of the high-frequency limit}\label{sec:reexamine_high}

We reexamine the high-frequency limit more carefully. In the text, we take the time average of $\varphi$ over the period $2\pi/\Omega$ so that the frequencies higher than $\Omega$ are cut off. In reality, the averaging does not provide a sharp high-frequency cutoff at $\Omega$ and has a ``tail" in the high frequency region. Namely, even $\slow$ has some fast components (we will explicitly see this below.) 

In order to assure the cutoff at $\Omega$, one would introduce an intermediate time scale $\tau_0$ such that $1/\Ts \gg \tau_0 \gg 2 \pi/\Omega$ and take the time average over the period $\tau_0$ instead of $2\pi/\Omega$. (In this Appendix, ``~$\bar{~}$~" denotes the time average over $\tau_0$.)

In order to estimate the cutoff effect, consider an arbitrary function $g(t)$. The Fourier transformation of $\overline{ g(t) }$ becomes
\begin{align}
\int^\infty_{-\infty} \frac{dt}{ \sqrt{2\, \pi} }~e^{i \omega t}\, 
  \overline{g(t)} 
& = \int^\infty_{-\infty} \frac{dt}{ \sqrt{2\, \pi} }~e^{i \omega t}\, 
  \frac{1}{\tau_0}\, \int_{t-\tau_0/2}^{t+\tau_0/2} dv\, g(v)
\\
& = \frac{ \sin(\omega\tau_0/2)}{ \omega\tau_0/2 }\, \tilde{g}(\omega)~.
\label{eq:filter}
\end{align}
The function $f(\omega) :=\sin(\omega\tau_0/2)/(\omega\tau_0/2)$ acts as a ``low-pass filter" with a high-frequency cutoff. But $f(\omega) =O(1)$ at $\omega\tau_0 \simeq 1$. If $\tau_0 = 2\pi/\Omega$, the components with $\omega \gtrsim \Omega$ could remain. By taking a large $\tau_0 \gg 2\pi/\Omega$, one can cut off these components.
 
Then, $\overline{ \oscA\slow }$ estimated in \eq{approx1} can be reestimated as
\be
\overline{ \oscA\slow } 
  = \frac{1}{\Omega\, \tau_0}\, O\big( |\, \oscA\, \slow\, | \big)
  \ll O\big( |\, \oscA\, \slow\, | \big)~,
%
\ee
where we used \eq{filter}. The difference from \eq{approx1} is that one does not even have to assume that $\slow$ is constant over $\tau_0$. The rest of the analysis is the same as the one in the text.

\end{document}